\documentclass[11pt, letterpaper]{article}

\usepackage{amsmath}
\usepackage{amssymb}
\usepackage{amsthm}
\usepackage{graphicx}
\usepackage{hyperref}
\usepackage{url}
\usepackage{algorithm}
\usepackage{algorithmic}
\usepackage{caption}
\usepackage{subcaption}
\usepackage{booktabs}
\usepackage{authblk}
\usepackage{appendix}
\usepackage[utf8]{inputenc}
\usepackage[T1]{fontenc}
\usepackage[english]{babel}
\usepackage{geometry}
\usepackage{xcolor}
\usepackage{fancyhdr}
\usepackage{microtype}
\usepackage{parskip}
\usepackage{tabularx}
\usepackage{enumitem}

\hypersetup{
    colorlinks=true,
    linkcolor=blue,
    filecolor=magenta,      
    urlcolor=cyan,
    citecolor=red,
}

\definecolor{headercolor}{RGB}{0, 50, 100}
\title{Exploring Multimodal Approaches for Alzheimer's Disease Detection Using Patient Speech Transcript and Audio Data}

\author[1]{Hongmin Cai}
\author[1]{Xiaoke Huang}
\author[2]{Zhengliang Liu}
\author[1]{Wenxiong Liao}
\author[2]{Haixing Dai}
\author[2]{Zihao Wu}
\author[3]{Dajiang Zhu}
\author[4]{Hui Ren}
\author[4]{Quanzheng Li}
\author[2]{Tianming Liu}
\author[4]{Xiang Li}

\affil[1]{South China University of Technology, Guangzhou, Guangdong, China \authorcr 
\{hmcai\}@scut.edu.cn \authorcr 
\{csxkhuang,cswxliao\}@mail.scut.edu.cn}
\affil[2]{University of Georgia, Athens, Georgia, USA \authorcr \{zl18864,haixing.dai,zihao.wu1,tliu\}@uga.edu}
\affil[3]{University of Texas at Arlington, Arlington, Texas, USA \authorcr \{dajiang.zhu\}@uta.edu}
\affil[4]{Massachusetts General Hospital, Boston, Massachusetts, USA \authorcr \{hren2,li.quanzheng,xli60\}@mgh.harvard.edu}

\date{}

\begin{document}

\maketitle

\begin{abstract}
Alzheimer's disease (AD) is a common form of dementia that severely impacts patient health. As AD impairs the patient's language understanding and expression ability, the speech of AD patients can serve as an indicator for this disease. This study investigates various methods for detecting AD using patient's speech and transcripts data from the DementiaBank Pitt database. The proposed approach involves pre-trained language models and Graph Neural Network (GNN) that constructs a graph from the speech transcript and extracts features using GNN for AD detection. Data augmentation techniques, including synonym replacement, GPT-based augmenter and so on, were used to address the small dataset size. Audio data was also introduced, and WavLM model was used to extract audio features. These features were then fused with text features using various methods. Finally, a contrastive learning approach was attempted by converting speech transcripts back to tts audio and using it for contrastive learning with the original audio. We conducted intensive experiments and analysis on the above methods. Our findings shed light on the challenges and potential solutions in AD detection using speech and audio data.
\end{abstract}

\section{Introduction}

Alzheimer's disease (AD), named after German psychiatrist Alois Alzheimer, is the most common form of dementia. AD degenerates brain cells, seriously affecting patients' quality of life \cite{breijyeh_comprehensive_2020}. Patients with AD have a shorter life expectancy with a median survival time of 3 to 6 years after diagnosis, which is even shorter for those with other underlying diseases \cite{helzner_survival_2008}. Memory loss is a hallmark symptom of AD, with patients getting lost and forgetting family and friends \cite{kumar_alzheimer_2022}. AD also impairs language ability, making it difficult to read, write, and communicate \cite{kumar_alzheimer_2022}, and may cause difficulty swallowing and movement disorders in advanced stages \cite{apostolova_alzheimer_2016}. AD is also a significant societal and economic burden, with 50 million people worldwide afflicted by the disease in 2020, projected to rise to 152 million by 2050 \cite{breijyeh_comprehensive_2020}. The cost of AD diagnosis, treatment, and care worldwide is estimated to be around 1 trillion US dollars per year \cite{breijyeh_comprehensive_2020}. 

While AD is incurable, early diagnosis can slow down its development, making it crucial to detect the disease at its early stages \cite{sperling_toward_2011}. However, medical diagnostic methods are often expensive, invasive, and require specialized equipment, necessitating the development of low-cost, non-intrusive, and simple diagnostic methods. Given that AD can impair patients' speech, their speech patterns exhibit certain characteristics, such as frequent silence, incoherence, word retrieval difficulty, and repetition \cite{liu_transfer_2022}. In this study, we utilized patient-generated speech and corresponding transcribed text data, applying various NLP-based methods to diagnose AD and compare their effectiveness. Through thorough analysis and comparison of various NLP-based methods, we aim to provide valuable insights and help advance the development of more effective diagnostic tools for Alzheimer's disease.

\section{Related Work}

There is prior work that utilizes patients' speech transcript for AD detection and diagnosis. A study by Ben Ammar et al. \cite{ben_ammar_speech_2018} proposed an AD detection model that extracts linguistic features from patient speech transcripts and performs feature selection based on the KNN algorithm. The selected features are then used to train a machine learning classifier, such as SVM, for the final diagnosis. Yamanki et al. \cite{yamanki_semantic_2022} proposed a contrastive learning model based on Siamese BERT to extract discriminative features from both the text of the patient speech transcripts and other features such as demographic, lexical, semantic information. The extracted features are then used for AD diagnosis using machine learning classifiers. The work by Roshanzamir et al. \cite{roshanzamir_transformer-based_2021} developed text data processing pipeline for analyzing patient speech transcript data. The pipeline consists of an augmentation module that enriches the input text data and a splitter that breaks text into sentences. The model then uses BERT to encode the sentences and the output encoded embeddings are used as input for various classification models, including multi-layer perceptron (MLP), convolutional neural network (CNN), and bidirectional LSTM (biLSTM). 

Some other work also incorporates speech data of patients to assist in AD diagnosis. A study proposed by Bertini et al. \cite{bertini_automatic_2022} uses log mel spectrograms and audio data augmentation techniques. The patients' audio data are first converted into log mel spectrograms, which is then enhanced using data augmentation techniques. Then, an autoencoder learns a condensed representation of the spectrogram, which then serves as input for a multilayer perceptron. Martinc et al. \cite{martinc_temporal_2021} use audio feature engineering for diagnosing AD. Specifically, they extract acoustic and textual features from the speech segments using openSMILE toolkit and GloVe, and further extract Active Data Representation (ADR) features based on them. 
A study by Agbavor et al. \cite{agbavor_artificial_2023} uses pre-trained audio model including data2vec and wav2vec2 to extract audio features from patients, and evaluates its performance on the ADReSSo (Alzheimer's Dementia Recognition through Spontaneous Speech only) dataset.

\section{Methods}

\begin{figure*}[t]
\centering
\includegraphics[width=0.95\textwidth]{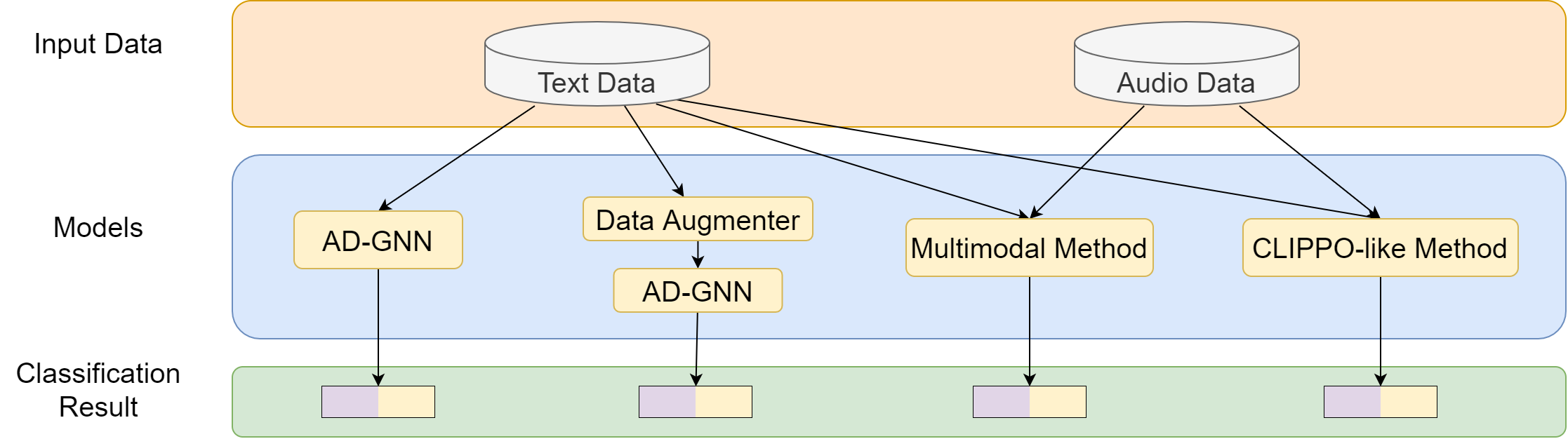}
\caption{Basic framework of our four methods.}
\label{methods_basic_framework}
\end{figure*}

In our work, we employ 4 approaches to diagnose AD using audio recordings and transcripts of audio. First, we attempt a GNN-based method, which we call AD-GNN. This method utilizes the pre-trained language model BERT and GNN to extract features from patient speech transcription and performs classification based on these features. Next, in the second approach, we first use text data augmenters to augment patient speech transcription and then use AD-GNN for classification. We hope that data augmentation can partially address the overfitting problem caused by small dataset size. In the third approach, we use both audio and text data, i.e., patient audio and its transcription. We expect the introduction of audio data can improve the performance. This approach employ pre-trained speech model to extract audio features and AD-GNN to extract text features, and we attempt various modal fusion methods. The fourth approach, called the CLIPPO-like method, also use both audio and text data. The innovation of this method lies in its imitation of CLIPPO's \cite{tschannen_clippo_2023} work. The text data is converted into TTS audio and then the same model is used to extract features from both the TTS audio and the original audio for contrastive learning. Fig. \ref{methods_basic_framework} illustrates the basic framework of these four methods.

\subsection{GNN-based Method}

There is a growing trend of developing and applying graph-based representation learning methods to better model complex structural patterns in real-world data. Graph Neural Network (GNN) has achieved great success in both computer vision \cite{zhou2020graph} and NLP \cite{goldberg2016primer}, with applications in the medical domain (e.g., the analysis of medical images \cite{guo2019predicting}, medical notes \cite{liu2022} and radiology reports \cite{jing2017automatic}). Thus, we envision that similar methods can also improve the effectiveness of NLP-based AD detection. At the same time, however, there is limited work related to detecting AD using graph-based methods.

In this work, we propose a lightweight, graph-based classification model for AD diagnosis using patient speech transcripts. The proposed model first computes the text embedding of the patient's speech with a pre-trained language model (BERT) \cite{devlin2018bert}, taking advantage of the pre-trained language model that has been trained on a massive text corpus using a large-scale model architecture. After this step, the proposed model constructs the graph representation of the embedding and utilizes Graph Neural Network (GNN) to learn discriminative features for the final disease classification. Our hypothesis is that the graph structure and the corresponding GNN-extracted features would provide more complex and context-rich representations, compared to representations learned from language models alone (which is the typical approach of existing work). The proposed AD-GNN model is evaluated on the patient speech transcripts data from the DementiaBank Pitt database \cite{becker_natural_1994}. We  formulate this problem as a binary (AD vs. normal) classification task and compare the AD-GNN model with previous methods.

\subsubsection{Notation}

The input of our model is a piece of text $\mathbf{T}=\left[\boldsymbol{t}_1, \boldsymbol{t}_2,\cdots,\boldsymbol{t}_n\right]$, which can be regarded as a token sequence with length $n$. $\mathbf{T}$ gets its initial embedding $\mathbf{H}^0=\left[\boldsymbol{h}_{1}^0, \boldsymbol{h}_{2}^0, \cdots, \boldsymbol{h}_{n}^0\right]$ through the embedding initializer where the initial embedding of token $\boldsymbol{t}_i$ is represented as $\boldsymbol{h}_{i}^0$. Then, we construct a graph $\mathbf{G}=\left(\mathbf{V},\mathbf{E}\right)$ according to $\mathbf{T}$ which consists of a set of $n$ nodes $\mathbf{V}=\left\{{v}_1,{v}_2,\cdots, {v}_n\right\}$ and edges $({v}_i,{v}_j)\in \mathbf{E}$. The adjacency matrix of $\mathbf{G}$ is denoted as $\mathbf{A}$. The feature of node ${v}_i$ is initialized as $\boldsymbol{h}_i^0$ and will be updated to $\boldsymbol{h}_i^K$ through a $K$-layer GNN network. $\mathbf{H}^K=\left[\boldsymbol{h}_{1}^K, \boldsymbol{h}_{2}^K, \cdots, \boldsymbol{h}_{n}^K\right]$, i.e., the final embedding of $\mathbf{T}$, will be fed into the classifier and the model outputs the final one-hot (binary classes) classification results.

\subsubsection{Model Architecture}

\begin{figure*}[t]
\centering
\includegraphics[width=0.95\textwidth]{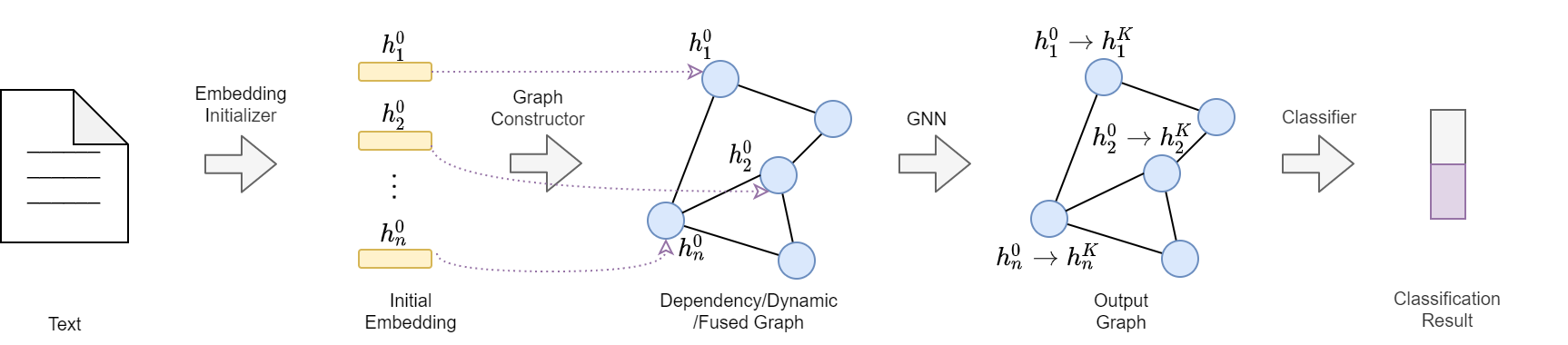}
\caption{The framework of AD-GNN.}
\label{gnn_model}
\end{figure*}

AD-GNN is based on the Graph4NLP project \cite{wu_graph_2021}, which enables efficient development and experimentation with GNN for NLP tasks such as text classification, semantic parsing, and machine translation. The algorithmic pipeline of AD-GNN is shown in Fig. \ref{gnn_model}. AD-GNN uses raw text as input and passes it to the embedding initializer to embed each token. Then, a graph constructor is adopted to construct the initial graph, where each node represents a token in the original text. After this step, the graph passes through GNN layers, and the node embeddings are updated. Finally, the classification result is obtained through the classifier. We will introduce the model details below.

\paragraph{Embedding Initialization}

Text $\mathbf{T}$ is a sequence of tokens. The first step of the AD-GNN model is to convert the tokens into initialized embeddings $\mathbf{H}^0$. Word vectors and pre-trained language models like BERT are widely used for embedding initialization. Graph4NLP library provides many strategies for token embeddings, and we chose two of them based on preliminary experiments:
\begin{itemize}
    \item \textbf{w2v\_bilstm}. We identify the word vector (glove.840B.300d.word2vec) for each token and then feed these representations to BiLSTM to obtain contextual information. The outputs of BiLSTM are the embeddings needed for further processing.
    \item \textbf{w2v\_bert\_bilstm}. First, we use word vectors for the initial embedding. Then, BERT is adopted to encode the input text. Finally, we concatenate the word vector embedding and the BERT embedding into one vector and use it as the input to BiLSTM to compute the final embedding.
\end{itemize}

\paragraph{Graph Construction}

Graph4NLP provides many strategies to build a graph from text sequences. Here we experiment with three of them: dependency graph, dynamic graph, and fused graph.

The first method is to build a dependency graph based on the dependency relationship between words to describe the text structure. Stanford CoreNLP \cite{chen_fast_2014} implements a transition-based dependency parser based on a neural network. It is worth noting that if the input text contains multiple sentences, after obtaining the dependency tree of each sentence, we will connect the last node of the dependency tree of the previous sentence with the header node of the next sentence with an edge to produce a connected graph.

The second method is to build a dynamic graph whose structure can evolve during the training process. The rationale is that static graphs (e.g., dependency graphs) may be incomplete or improper, and the errors produced in the graph construction phase cannot be corrected. These errors may affect the accuracy of the final classification results. To counter this problem, Chen et al. proposed an end-to-end dynamic graph learning framework, Iterative Deep Graph Learning (IDGL) \cite{chen_iterative_2020}, which is used in AD-GNN. IDGL first calculates the similarity between every two nodes to obtain a complete graph and then performs graph sparsification to generate a sparse graph. We use weighted cosine similarity to measure the similarity between nodes. 
\begin{equation}
    \mathbf{S}_{i j}=\cos \left(\mathbf{W} \odot \boldsymbol{h}_i^0, \mathbf{W} \odot \boldsymbol{h}_j^0\right), \label{weighted_cosine_sim}
\end{equation}
where $\mathbf{S}_{i j}$ denotes weighted cosine similarity between node ${v}_i$ and node ${v}_j$, $\mathbf{W}$ denotes a trainable weight vector, $\odot$ denotes the Hadamard product, and $\boldsymbol{h}_i^0$ and $\boldsymbol{h}_j^0$ are the embeddings of node ${v}_i$ and ${v}_j$, respectively. There are various graph sparsification techniques in IDGL and $\varepsilon$-neighborhood method is adopted in AD-GNN, because it only retains the connection with a weight greater than a pre-defined threshold $\varepsilon$.
\begin{equation}
    \mathbf{A}_{i, j} = \left\{
    \begin{array}{ll}
        \mathbf{S}_{i, j} & \mathbf{S}_{i, j}>\varepsilon \\
        0 & \text{otherwise}
    \end{array} \right. , \label{epsilon_neighborhood}
\end{equation}
where $\mathbf{S}_{i, j}$ is the similarity between node ${v}_i$ and ${v}_j$, $\mathbf{A}_{i, j}$ is the weight of edge between node ${v}_i$ and ${v}_j$ in the sparse graph.

Another method for graph construction is to fuse dependency and dynamic graph together to form a new graph. The fusion method implemented by Graph4NLP can be represented as
\begin{equation}
    \mathbf{A}_\text{com} = \lambda \mathbf{L}_\text{dep} + (1-\lambda)\operatorname{f}(\mathbf{A}), \label{combine_static_and_dynamic_graph}
\end{equation}
where $\mathbf{A}_\text{com}$ is the adjacency matrix of the new graph, $\lambda$ is a hyperparameter with a value between 0 and 1, $\mathbf{L}_\text{dep}$ is the normalized Laplacian matrix of the dependency graph, $\mathbf{A}$ is the adjacency matrix of the dynamic graph and $\operatorname{f}(\cdot)$ is the matrix normalization operation (e.g., row normalization).

\paragraph{GNN Layers}
The initialized graph is fed into GNN layers to learn the feature representation of each node. In the proposed AD-GNN model, GraphSAGE (Graph SAmple and aggreGatE) \cite{hamilton_inductive_2018} and GGNN (Gated Graph Neural Network) \cite{li_gated_2017} with $K$ layers are separately tested. For the GraphSAGE model, its forward propagation operation of node $v$ in layer $k$ consists of three steps: 

\begin{enumerate}
    \item Considering the calculation efficiency, the random sampling method is adopted to sample $n$ neighbors of each node $v$ in graph, which are denoted by $\mathcal {N} (v)$
    \item Aggregate the embedding of neighbor nodes $\mathcal {N} (v)$ through the aggregation function $\operatorname{AGGREGATE}_{k}\left(\cdot\right)$ 
 (usually LSTM) to obtain the embedding of $\mathcal {N} (v)$, i.e.
 \begin{equation}
     \boldsymbol{h}_{\mathcal{N}(v)}^{k} = \operatorname{AGGREGATE}_{k}\left(\left\{\boldsymbol{h}_{u}^{k-1}, \forall u \in \mathcal{N}(v)\right\}\right).
 \end{equation}
    \item Concatenate the embedding of node $v$ at $k-1$ layer and the embedding of neighbors of node $v$ at $k$ layer, and pass the result through a full connection layer to get the embedding of node $v$ at layer $k$, i.e.
    \begin{equation}
        \boldsymbol{h}_{v}^{k} = \sigma\left(\mathbf{W}^{k} \cdot \operatorname{CONCAT}\left(\boldsymbol{h}_{v}^{k-1}, \boldsymbol{h}_{\mathcal{N}(v)}^{k}\right)\right),
    \end{equation}
    where $\mathbf {W} ^ k$ is a trainable matrix and $\sigma$ is the activation function.
\end{enumerate}

For the GGNN model, which is a GRU-based message passing model, its forward propagation operation of node $v$ in layer $k$ consists of three steps:

\begin{enumerate}
    \item The first step is to perform message passing operation, that is, node $v$ and its adjacent nodes interact through edges. The result of message passing operation $\boldsymbol{a}_v^{k}$ is 
    \begin{equation}
        \boldsymbol{a}_v^{k}=\mathbf{A}_{v:}^{\top}\left[{\boldsymbol{h}_1^{k-1}}^{\top} \ldots {\boldsymbol{h}_{n}^{k-1}}^{\top} \right]^{\top}+\boldsymbol{b},
    \end{equation}
    where $\mathbf{A}_{v:}^{\top}$ denotes a matrix where the element in row $i$ and column $0$ represents the feature of the edge from node $i$ to node $v$, and the element in row $i$ and column $1$ represents the feature of the edge from node $v$ to node $i$.
    \item The second step is to generate new information. $\boldsymbol{r}_v^k$ is the reset gate that controls the generation of new information. 
    \begin{equation}
        \boldsymbol{r}_v^k =\sigma\left(\mathbf{W}^r \boldsymbol{a}_v^{k}+\mathbf{U}^r \boldsymbol{h}_v^{k-1}\right),
    \end{equation}
    where $\mathbf{W}^r$ and $\mathbf{U}^r$ are both trainable matrices. The resulting new information $\widetilde{\boldsymbol{h}_v^{k}}$ is
    \begin{equation}
        \widetilde{\boldsymbol{h}_v^{k}} =\tanh \left(\mathbf{W} \boldsymbol{a}_v^{k}+\mathbf{U}\left(\boldsymbol{r}_v^k \odot \boldsymbol{h}_v^{k-1}\right)\right),
    \end{equation}
    where $\mathbf{W}$ and $\mathbf{U}$ are both trainable matrices and $\odot$ denotes Hadamard product.
    \item The last step is to forget certain old information and remember new information to obtain the final representation. The update gate $\boldsymbol{z}_v^k$ controls what information should be remembered.
    \begin{equation}
        \boldsymbol{z}_v^k =\sigma\left(\mathbf{W}^z \boldsymbol{a}_v^{k}+\mathbf{U}^z \boldsymbol{h}_v^{k-1}\right),
    \end{equation}
    where $\mathbf{W}^z$ and $\mathbf{U}^z$ are both trainable matrices. The final feature of node $v$ at layer $k$ is calculated as
    \begin{equation}
        \boldsymbol{h}_v^{k} =\left(\boldsymbol{1}-\boldsymbol{z}_v^k\right) \odot \boldsymbol{h}_v^{k-1}+\boldsymbol{z}_v^k \odot \widetilde{\boldsymbol{h}_v^{k}}.
    \end{equation}
\end{enumerate}

\paragraph{Classifier}
AD-GNN uses a pooling layer and a multi-layer perceptron (MLP) to perform the final classification. An average pooling layer is used to average the features across all nodes to characterize the whole graph, i.e.,
\begin{equation}
    \boldsymbol{r} = \frac{1}{n}\sum_{i=1}^{n} \boldsymbol{h}_i^K,
\end{equation}
where $\boldsymbol{h}_i^K$ denotes the final feature of node $v_i$, and $\boldsymbol{r}$ denotes the feature of the whole graph. The MLP layer accept the averaged graph feature as input and performs the binary classification.

\subsubsection{Loss Function}
The loss function of the proposed AD-GNN model varies according to the graph construction method. For the dependency graph, the loss is simply the cross entropy loss $L_{\text{pred}}$ of the classification. For dynamic graph or fused graph, the smoothness, connectivity and sparsity are considered for the regularization of the constructed graph:
\begin{equation}
    \begin{aligned}
        L_{\text{G}}= &\frac{\alpha}{n^2} \operatorname{tr}\left(\left(\mathbf{H}^0\right)^T \mathbf{L} \mathbf{H}^0\right) \\ 
        &+ \frac{-\beta}{n} \mathbf{1}^T \log (\mathbf{A} \mathbf{1}) \\
        &+ \frac{\gamma}{n^2}\|\mathbf{A}\|_F^2
    \end{aligned}, \label{regularization_loss}
\end{equation}
where $\alpha$, $\beta$, $\gamma$ are hyperparameters, $n$ is the number of nodes, $\operatorname{tr}\left(\cdot\right)$ denotes the trace of the matrix, $\mathbf{H}^0$ is the node feature matrix, $\mathbf{A}$ is the adjacency matrix, $\mathbf{L}$ is the unnormalized graph Laplacian matrix, $\mathbf{1}$ denotes a vector whose elements are all one and $\|\cdot\|_F$ denotes the F-norm of a matrix. The first term, the second term, the third term penalizes non-smoothness, disconnection, and the density of the graph, respectively. The final loss is defined as the sum of classification loss and graph regularization loss, that is,
\begin{equation}
   L=L_{\text{pred}}+L_{\text{G}} .
\end{equation}

\subsection{Data Augmentation Method}

Data augmentation is a method that can expand the variety of data used to train models by generating modified versions of existing data. In order to address the issue of limited data in the Pitt Cookie-Theft dataset, we use a variety of text data augmentation techniques. The goal of these methods is to artificially increase the size of our dataset, thereby providing our models with more diverse training examples, reducing the risk of overfitting, and potentially improving their performance. Here is an introduction to the data augmentation methods we use.

\begin{itemize}
    \item \textbf{Synonym Replacement.} This method uses a synonym dictionary, such as WordNet \cite{miller1995wordnet}, to replace words in the original text with their synonyms. This method aims to maintain the semantic integrity of the sentences while introducing lexical diversity.
    \item \textbf{Counter-fitting Embedding Replacement.} The data augmentation method based on word embedding \cite{wang2015s} refers to replacing words in a sentence with other words that are close to them in the embedding space. However, this poses a problem as two words with similar embeddings only indicate that they typically occur in similar contexts, but their semantics may not necessarily be the same and may even be opposite. The counter-fitting embedding data augmentation method \cite{alzantot_generating_2018} uses counter-fitting embedding instead, which reduces the distance between synonyms and increases the distance between antonyms, thus better ensuring semantic consistency.
    \item \textbf{Masked Language Model Augmentation.} This method \cite{kobayashi_contextual_2018,li_contextualized_2021} leverages the power of pre-trained language models such as RoBERTa model \cite{liu2019roberta} to generate new sentences. It uses the RoBERTa model to perform three operations: token swap, token insert, and token merge. Specifically, token swap operation means randomly replace tokens in the original text with a special "[MASK]" token. Token insert operation means randomly insert "[MASK]" tokens into the original text. Token merge operation means randomly merged adjacent tokens into a single "[MASK]" token. Subsequently, it uses RoBERTa to predict the most likely token for the masked positions.
    \item \textbf{Random Sentence Deletion.} This method randomly removes sentences from the original text. This technique aims to simulate missing or incomplete information, which is common in real-world scenarios. This process helps to improve the model's robustness and generalization ability by training it to make accurate predictions even when confronted with incomplete data.
    \item \textbf{Augmentation with GPT Models.} This method \cite{dai_auggpt_2023} uses ChatGPT to rephrase each samples. By using appropriate prompt, chatGPT can rephrase the patient's speech transcript data as instructed and thus effectively increase the size of training set.
\end{itemize}

\subsection{Multimodal Method}

\begin{figure*}[t]
\centering
\includegraphics[width=0.95\textwidth]{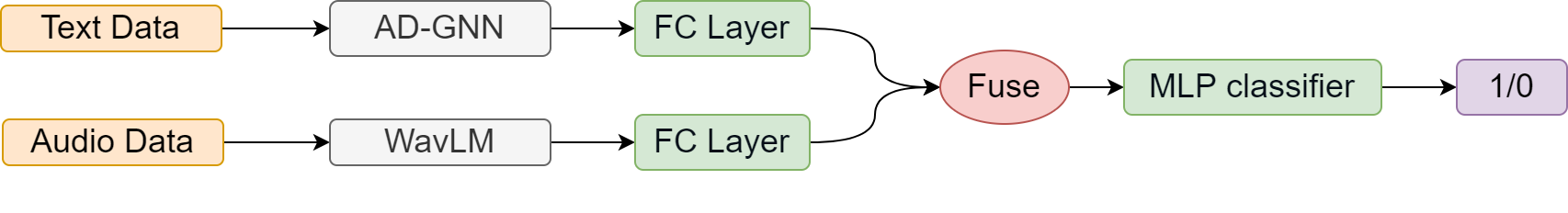}
\caption{The framework of our multimodal method.}
\label{multimodal_model}
\end{figure*}

Audio data can provide additional information that is not captured in transcriptions. For instance, changes in speech patterns, such as pace, tone and rhythm, especially unsmooth speech, such as stuttering and pauses, are often early indicators of cognitive decline in AD patients. These features can be extracted from audio data but are lost in text transcriptions.

In this work, we use WavLM \cite{chen_wavlm_2022} model to extract audio features and perform classification. WavLM is a universal pre-trained speech model developed by Microsoft. The model initially applies random transformations to the input speech signal, such as mixing and adding noise, to enhance the model's robustness. Subsequently, a CNN encoder and a Transformer encoder are used to extract audio features. The Transformer encoder employs gated relative position bias to better capture the sequence order of the input speech. Furthermore, WavLM adopts a masked loss similar to BERT. Specifically, it uses the K-means algorithm to convert speech features into discrete labels, and predicts the labels of masked positions. WavLM is pre-trained on large-scale and diverse speech data, including e-books, YouTube videos, European Parliament recordings and so on, totaling 94,000 hours. Therefore, compared to traditional audio feature extraction methods, it exhibits excellent robustness and generalization capabilities.

Next, we employ multimodal learning. Multimodal learning, which combines information from different types of data (in this case, text and audio), can lead to more robust and accurate models. This is because different modalities can provide complementary information, allowing the model to learn from a more generalizable representation of the data. We pass the text features obtained by AD-GNN and the audio features extracted by WavLM through a fully connected layer, respectively, to ensure that the text and audio features have the same dimension. Then we fuse these two features to facilitate sufficient interaction between them. The fused features are then fed into the MLP classifier for classification, yielding the final result. Our multimodal approach is illustrated in Fig. \ref{multimodal_model}.

The fusion method is crucial for the effectiveness of multimodal models. We have attempted two multimodal fusion methods: direct concatenation and cross network. Direct concatenation is the simplest fusion method, which directly concatenates the audio and text features into one tensor. Cross network proposed by Wang et al. \cite{wang_deep_2017} can explicitly apply feature crossing. It consists of multiple layers, where each layer produces higher-order interactions based on existing ones, and keeps the interactions from previous layers.

\subsection{CLIPPO-like Method}

\begin{figure*}[t]
\centering
\includegraphics[width=0.95\textwidth]{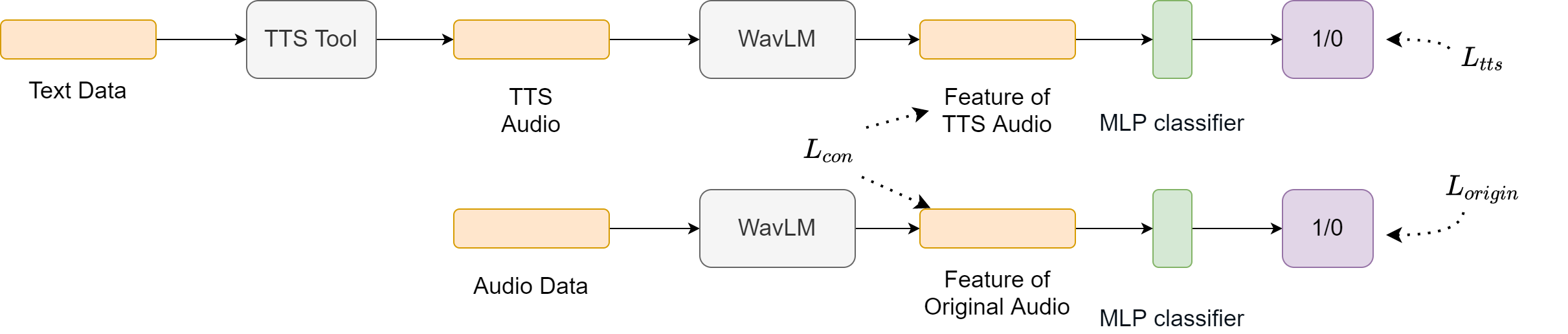}
\caption{The framework of our CLIPPO-like method.}
\label{clippo_like_model}
\end{figure*}

CLIPPO (CLIP-Pixels Only) \cite{tschannen_clippo_2023} is a pixel-based multimodal model that can understand both images and alt-text simultaneously without requiring text encoders or tokenizers. Its approach is to render alt-text as images and then encode both images and text using the same network architecture. It uses a contrastive learning loss function to make embeddings of matching images and alt-text as close as possible and all other image and alt-text embeddings as far apart as possible. The importance of CLIPPO lies in its achievement of unified modeling of images and text, not limited by text tokenizers, simplifying the complexity of multimodal learning, and improving model scalability and generalization. Compared to using two completely different models for text and image modalities, CLIPPO has reduced the number of parameters by half while achieving comparable experimental results on various tasks, including zero-shot image classification and image-text retrieval and so on.

In our research, we attempt to replicate the work of CLIPPO. Fig. \ref{clippo_like_model} illustrates our model architecture. We follow these steps:

\begin{enumerate}
    \item We use SpeechT5 model \cite{ao_speecht5_2022} fine-tuned for text-to-speech (TTS) to convert the patients' transcripts back into speech. The speech features of these TTS audios, such as intonation and speed, are similar to those of normal human speech.
    \item Next, we use WavLM to extract features from both the original audio and the TTS audio.
    \item We then employ contrastive learning to compare the generated TTS audio with the patient's original speech data. Similar to CLIPPO, we aim to make the features of the matched original and TTS audio as similar as possible, while keeping the features of other original and TTS audio pairs as different as possible. This approach can help the model extract richer and more discriminative features. We use a contrastive loss function similar to that used in CLIPPO. Specifically, in each batch, there are $n$ samples, and the original audio feature of the $i$-th sample is represented as $a_i^{origin}$, while the TTS audio feature of the $i$-th sample is represented as $a_i^{tts}$. First, we perform L2 normalization on the above features to obtain $e_i^{origin}, i={1,2,\cdots,n}$ and $e_i^{tts}, i={1,2,\cdots,n}$. Then, we calculate the cosine similarity matrix $S$ between the original audio features and TTS audio features. $S$ is an $n\times n$ matrix, and the element in the $i$-th row and $j$-th column $S_{ij}$ is calculated as $S_{ij}=e_i^{origin}e_j^{tts}e^{t}$, where $t$ is a temperature parameter. Among these $n\times n$ sample pairs, there are $n$ positive sample pairs (matching original audio and TTS audio pairs) and $n^2-n$ negative sample pairs (all other original audio and TTS audio pairs). We separately calculate the cross-entropy loss of original audio $L_{con}^{origin}$ and the cross-entropy loss of TTS audio $L_{con}^{tts}$. Finally, we take the average of these two loss as the final contrastive loss $L_{con}$. That is:
    \begin{equation}
        \begin{aligned}
            L_{con}^{origin} &= \operatorname{cross\_entropy\_loss}(S, labels, axis=0),\\
            L_{con}^{tts} &= \operatorname{cross\_entropy\_loss}(S, labels, axis=1),\\
            L_{con} &= \frac{1}{2} \left(L_{con}^{origin}+L_{con}^{tts}\right), \\
        \end{aligned}
    \end{equation}
    where $labels=[1,2,\cdots,n]$.
    \item At the same time, we record the classification loss $L_{origin}$ for the original audio and the classification loss $L_{tts}$ for the TTS audio, both of which are calculated using cross-entropy loss.
    \item The final loss is a weighted sum of the above three losses, namely, 
    \begin{equation}
        L=\alpha L_{con} + \beta L_{origin} + \gamma L_{tts}. \label{loss_of_clippo_like_method}
    \end{equation}
\end{enumerate}

The advantage of this approach lies in the fact that since the TTS audio is directly generated from text, it contains semantic information that may not be well reflected in the original speech. By using contrastive learning, we can make the features of the original audio and its corresponding TTS audio as similar as possible. This allows our model to learn how to link speech features such as intonation and pauses with textual features such as semantic information, without the need for pre-trained language models such as BERT. This approach reduces the number of parameters required, resulting in a more compact model structure.

\section{Experiments}

\subsection{Datasets}

In this work, we used patients' speech and corresponding transcript data from the DementiaBank Pitt database \cite{becker_natural_1994}. Participants were asked to describe what they saw when presented with a picture showing a mother washing dishes in a sink and children trying to steal cookies from a cookie jar (the "Cookie-Theft picture"). The dataset consists of 257 samples marked "probable/possible AD" and 243 samples marked "healthy control." The audio records of these interviews were manually transcribed and annotated. The transcriptions are in the CHAT (Codes for the Human Analysis of Transcripts) format \cite{macwhinney2000childes}, a standard protocol for TalkBank data.

\subsection{Experimental Settings}
In the experiments part, we have conducted extensive experiments on GNN-based method, data augmentation-based method, multimodal method, and CLIPPO-like method. For GNN-based method, we separately test the effects of different choices of embedding initializer, graph constructor, and GNN on model performance. For the embedding initializer, we tested w2v\_bilstm and w2v\_bert\_bilstm. For graph construction methods, we compared dependency graph, dynamic graph, and the fusion of dependency graph and dynamic graph. For the choice of GNNs, we tested GraphSAGE and GGNN. We also experimented with adjusting the number of layers in the GNN, as well as completely removing the GNN for ablation study. The batch size, epochs, and learning rate are set as 20, 30, and 0.001 respectively. The $\varepsilon$ in Eq. \ref{epsilon_neighborhood} is set as 0.95. The $\alpha,\beta,\gamma$ in Eq. \ref{regularization_loss} are set as 0.1, 0.1, and 0.3, respectively. The $\lambda$ in Eq. \ref{combine_static_and_dynamic_graph} is set as 0.5. The length of embedding $\boldsymbol{h}$ is set to 300.

For data augmentation methods, we first use various text data augmentation methods to increase the size of the training set, Then, just like in the previous experiment, we use AD-GNN for classification. The data augmentation methods we used are listed below.

\begin{itemize}
    \item \textbf{Counter-Fitting Embedding Augmenter.} This method replaces words in a paragraph with other words that are top-k similar to them in the counter-fitting embedding space.
    \item \textbf{Sentence Deletion Augmenter.} This method randomly removes one sentence from the original text.
    \item \textbf{RoBERTa Augmenter.} This method uses the RoBERTa model to perform three operations: token swap, token insert, and token merge.
    \item \textbf{Wordnet Augmenter.} This method substitutes words in the original text with their synonyms from the WordNet lexical database.
    \item \textbf{GPT-3.5 Augmenter.} This method uses the gpt-3.5-turbo model to generate new samples. The prompt we provided to the model is “Please write another paragraph using the speaking style of the following paragraph.”
    \item \textbf{GPT-4 Augmenter.} This method uses the gpt-4 model to generate new samples. We provide the same prompt to the model as used in GPT-3.5 Augmenter.
\end{itemize}

For all data augmentation methods except GPT-4 Augmenter, we attempted to double the size of the dataset by generating one new sample for each original sample. For GPT-4 Augmenter, due to cost considerations, we randomly selected one sample every five samples for data augmentation. Later, to investigate the impact of augmentation factor on performance, we also conducted experiment on the Sentence Deletion Augmenter with augmentation factor of 5 (that is, generating 5 new samples for each original sample). For data augmentation methods other than that based on GPT-3.5 and GPT-4, we use the implementation provided in the TextAttack \cite{morris2020textattack} library.

For the experiments on multimodal methods, we compared the effectiveness of using only AD-GNN for classifying transcriptions of patient speech, only using the WavLM-base model for classifying patient speech audio, and combining both with the multimodal method. For the multimodal method, we compared the effectiveness of two fusion methods: direct concatenation and cross network. The batch size, epochs, and learning rate are set as 20, 30, and 0.001, respectively, which is the same as in the GNN-based method experiment. The sampling rate of the audio data is 16,000. For the experiment of CLIPPO-like method, the batch size, learning rate, and epoch were set to 4, 1.5e-5, and 100, respectively. The parameters $\alpha$, $\beta$, and $\gamma$ in Eq. \ref{loss_of_clippo_like_method} were set to 1. 

For all the experiments mentioned above, we employed 10-fold cross-validations for 5 times to get stable results. We used accuracy for the evaluation metrics as the number of positive and negative samples is largely balanced. 

\subsection{Results}

\begin{table*}[t]
\centering
\resizebox{.95\columnwidth}{!}{
\begin{tabular}{c|c|c|c}
    \hline
    Embedding Initializer & Graph Construction Method & GNN & Accuracy \\
    \hline
    w2v\_bilstm & Dependency graph & One-layer GraphSAGE & $0.8088\pm 0.0556$ \\
    w2v\_bert\_bilstm & Dependency graph & One-layer GraphSAGE & $0.8460\pm 0.0492$ \\
    w2v\_bert\_bilstm & Dynamic graph & One-layer GraphSAGE & $0.8444\pm 0.0455$ \\
    w2v\_bert\_bilstm & Fused graph & One-layer GraphSAGE & $0.8460\pm 0.0420$ \\ 
    w2v\_bert\_bilstm & Fused graph & Two-layer GraphSAGE & $\textbf{0.8504}\pm 0.0517$ \\
    w2v\_bert\_bilstm & Fused graph & Three-layer GraphSAGE & $0.8492\pm 0.0462$ \\
    w2v\_bert\_bilstm & Fused graph & One-layer GGNN & $0.8444\pm 0.0420$ \\
    w2v\_bert\_bilstm & Fused graph & No GNN & $0.8484\pm 0.0509$ \\
    \hline
\end{tabular}}
\caption{Classification performance of AD-GNN using different combinations of the embedding algorithm, graph construction method, and GNN network.}
\label{table_results_of_gnn_model}
\end{table*}

The results related to AD-GNN are shown in Table \ref{table_results_of_gnn_model}. Apparently, w2v\_bert\_bilstm leads to better classification performance (compared to w2v\_bilstm), indicating that the pre-trained language model is more effective in performing the initial token embedding. However, we noticed that regardless of the graph construction method and GNN used, there was no significant improvement in accuracy, except for a slight improvement in accuracy, reaching 0.8504, when using the fused graph and two-layer GraphSAGE. The reason may be that the graph did not capture the key information related to AD detection. For example, dependency graph mainly focus on the grammatical relationship between words in a sentence, and these relationships may not be relevant to the language features of AD patients, such as vocabulary richness.

\begin{table*}[t]
\centering
\begin{tabular}{c|c|c}
    \hline
    Augmentation Method & Augmentation Factor & Accuracy \\
    \hline
    No Data Augmentation & N/A & $0.8460\pm 0.0420$ \\
    Counter-Fitting Embedding Augmenter & 1 & $0.8476 \pm 0.0551$ \\
    Sentence Deletion Augmenter & 1 & $0.8444\pm 0.0490$ \\
    RoBERTa Augmenter & 1 & $0.8408\pm 0.0511$ \\
    Wordnet Augmenter & 1 & $\textbf{0.8484} \pm 0.0541$ \\
    GPT-3.5 Augmenter & 1 & $0.8416\pm 0.0535$ \\
    GPT-4 Augmenter & 0.2 & $0.8476\pm 0.0503$ \\
    Sentence Deletion Augmenter & 5 & $\textbf{0.8484} \pm 0.0465$ \\
    \hline
\end{tabular}
\caption{Classification performance of AD-GNN using different text augmentation methods.}
\label{table_results_of_augmentation_method}
\end{table*}

The experimental results about data augmentation are shown in Table \ref{table_results_of_augmentation_method}. It is shown that all data augmentation methods have a negligible impact on the accuracy of the model. Among them, Counter-Fitting Embedding Augmenter, Wordnet Augmenter, GPT-4 Augmenter, and Sentence Deletion Augmenter (with augmentation factor of 5) can improve the performance, but the effect is not significant. Moreover, increasing the augmentation factor also has a very limited impact on the effect. This may be due to the following reasons. First, AD patients are used to using simple words, and synonym replacement may replace words with uncommon or complex ones. Moreover, although Sentence Deletion Augmenter can simulate the information loss in the real world, it may also cause the loss of important information. Additionally, excessive augmentation may introduce too much noise, making it difficult for the model to learn useful information. What's more, some data augmentation methods such as RoBERTa Augmenter may not preserve the semantics of the original sentence. In addition, data augmentation methods may not simulate the speaking style of AD patients. For example, for GPT3.5 Augmenter, even when we ask it to mimic the speaking style of the speaker in the prompt, its effect is sometimes unsatisfactory, and the generated sentences are too formal.

\begin{table*}[t]
\centering
\begin{tabular}{c|c|c}
    \hline
    Modal & Method & Accuracy \\
    \hline
    Audio & WavLM & $0.7714\pm 0.0538$ \\
    Text and Audio & Multimodal Method (Direct Concatenation) & $0.8475 \pm 0.0496$ \\
    Text and Audio & Multimodal Method (Cross Network) & $0.8418\pm 0.0528$ \\
    Text and Audio & CLIPPO-like Method & $\textbf{0.8484} \pm 0.0544$ \\
    \hline
\end{tabular}
\caption{Comparison of experimental results using only textual data, only audio data, and both textual and audio data.}
\label{table_results_of_multimodal_method}
\end{table*}

The experimental results about multimodal methods are shown in Table \ref{table_results_of_multimodal_method}. It can be observed that the accuracy of the text modality (0.8460) is much higher than that of the audio modality (0.7714). This may be due to the complexity and dimension of audio data are often higher than those of text data, which may make it difficult for the model to learn effective feature representations during the training process. In addition, background noise in audio may also affect the accuracy of the experiment. The reason might also be that although the WavLM model performs well on many speech tasks, it is only pre-trained on data such as electronic books and European Parliament recordings, and has not been pre-trained on AD patients' data. Therefore, it cannot fully capture audio features related to AD.

When we fused the data from both text and audio modalities, we found that the accuracy of the direct concatenation method (0.8475) is almost the same as that of the cross network method (0.8418), and is very close to the experimental result of using only text. Perhaps this is because the experimental results of the two modalities differ greatly, with strong performance in the text modality but mediocre results in the audio modality, causing the model to overly rely on the stronger modality.

But in the end, We found that the CLIPPO-like method significantly improves performance compared to using only raw audio. This indicates that aligning TTS voice with raw audio helps the model understand the semantic information of the audio without the need for pre-trained speech models like BERT.

\section{Conclusion and Discussion}

In this study, we systematically explored various methods for detecting Alzheimer's disease (AD) using patients' speech and their transcribed text data, including GNN-based methods, data augmentation methods, multimodal-based methods and CLIPPO-like methods, and conducted extensive experiments, which provided rich references for future research. Our study also conducted in-depth analysis which provided directions for future work. We also found that the CLIPPO-like method can enable the model to learn semantic information without introducing pre-trained language model, which can significantly improve the performance compared to using only raw audio.

For future research, there are several possible directions for improvement. First, in addition to using patient speech and their transcribed text, we can also combine other data, such as the patient's facial expressions, which may provide more information and improve the detection of AD. Second, we believe that bigger datasets may be the best way to improve model performance. In our study, we only used about 500 samples, which may have limited the performance of our models. Finally, we believe that improving data augmentation methods may be a promising direction. In our study, we tried various data augmentation methods, but the effects were not ideal. We believe that if the new samples generated by augmenter can better simulate the language features of AD patients, the model may have better performance.

\section{Reproducibility Statement}
\begin{itemize}
  \item \textbf{Datasets}: The patient audio and transcript data used in this work is provided by the DementiaBank Pitt database at dementia.talkbank.org \cite{becker_natural_1994}. Data will be available for download upon request. 
  \item \textbf{Code}: Both the code and experiment settings for our model are available at: \url{https://github.com/shui-dun/multimodal\_ad}.
\end{itemize}

\bibliography{LLM_refs}
\bibliographystyle{unsrt}

\end{document}